\PassOptionsToPackage{table,xcdraw}{xcolor}
\documentclass{gradient}
\usepackage{natbib}
\setcitestyle{authoryear,round,citesep={;},aysep={,},yysep={;}}

\usepackage{amsmath,amsfonts,bm}

\def\eqref#1{equation~\ref{#1}}

\def\1{\bm{1}}

\DeclareMathAlphabet{\mathsfit}{\encodingdefault}{\sfdefault}{m}{sl}
\SetMathAlphabet{\mathsfit}{bold}{\encodingdefault}{\sfdefault}{bx}{n}

\usepackage{array}
\usepackage{enumitem}
\usepackage{multirow}
\usepackage{listings}
\usepackage{xspace}
\usepackage{xurl}
\usepackage{mathpazo}
\usepackage{tikz}

\hypersetup{
  pdftitle={Zero2Repo: Can Coding Agents Build Repositories from Scratch?},
  pdfauthor={Pei Yang et al.},
  pdfsubject={Coding-agent evaluation and repository construction},
  pdfkeywords={coding agents, code generation, software engineering, benchmark}
}

\newif\ifdraft
\draftfalse
\ifdraft
  \newcommand{\todo}[1]{\textcolor{red}{[TODO: #1]}}
\else
  \newcommand{\todo}[1]{}
  
\fi

\newcommand{\bench}{Zero2Repo\xspace}
\newcommand{\cbrun}{\texttt{cbrun}\xspace}
\newcommand{\prd}{PRD\xspace}
\newcommand{\contract}{Interface Contract\xspace}

\title{Zero2Repo: Can Coding Agents Build Repositories from Scratch?}
\renewcommand{\shorttitle}{Zero2Repo}

\author{%
\small
Pei Yang$^{1}$, Tianyu Shi$^{2,\dagger}$, Yuhang Yao$^{3}$, Wanyi Chen$^{9}$, Tongyun Yang$^{9}$\\
Dun Pei$^{9}$, Haonan Wang$^{9}$, Pengbin Feng$^{4}$, Guanxu Yu$^{9}$, Jingchun Huang$^{9}$\\
Zeyu Zhang$^{9}$, Shuhan Sun$^{9}$, Hao Li$^{5}$, Alex Gu$^{6}$\\
Xiang Li$^{7}$, Jie Xiao$^{1}$, Xinyu Wang$^{2}$, Hanxin Chen$^{8}$, Daqi Li$^{9}$\\
Qi Jia$^{9}$, Hongshan Lin$^{9}$, Zhizhou Gu$^{9}$, Zijun Tian$^{9}$, Weizhi Du$^{9}$\\
Lynn Ai$^{1}$, Eric Yang$^{1}$
}

\affiliation{%
\scriptsize
$^{1}$Gradient Data \quad $^{2}$McGill University \quad
$^{3}$Carnegie Mellon University \quad $^{4}$University of Southern California\\[0.15em]
$^{5}$Queen's University \quad $^{6}$MIT \quad
$^{7}$University College London, University of London\\[0.15em]
$^{8}$University of California, San Diego \quad $^{9}$Independent Researcher\\[0.35em]
$^{\dagger}$Corresponding author: \href{mailto:tianyu.shi3@mcgill.ca}{tianyu.shi3@mcgill.ca}
}

\date{October 2026}
\sourcecode{https://github.com/Gradient-Data/Zero2Repo}
\homepage{https://zero2repo.ai/}

\begin{document}

\abstract{%
Coding agents are increasingly asked to build software rather than patch it, yet benchmarks for from-scratch repository construction are mostly limited to a single language and depend on manually curated tasks. We introduce \bench, a benchmark in which an agent receives a product requirements document, an interface contract, and an empty workspace, and must deliver a complete repository in the project's native ecosystem. Tasks are produced by a language-agnostic authoring pipeline that converts real, version-pinned open-source projects into behavioral specifications, reproducible environments, and hidden acceptance tests. Each task is validated by execution: a reference implementation derived from the upstream project must pass, and adversarial validation must show that the tests reject incorrect implementations. Evaluation runs production coding agents in isolated containers, withholds the acceptance tests until an explicit submission, and assigns a binary reward only when every test passes, with no LLM judge. The pipeline and harness make no language-specific assumptions and apply to mainstream programming ecosystems; the current release contains 17 tasks in Python, TypeScript, Go, C++, C, and Zig. Across 18 model--harness configurations and one adopted high-effort run per task, the strongest configuration solves 15 of 17. Of the 121 judged failures, 95 still pass at least 90\% of the hidden tests, while 51 of the 172 unsuccessful runs never receive a complete judge result. No configuration solves every task, and one task is solved by none.
}

\maketitle

\section{Introduction}
\label{sec:intro}

Writing a correct function and delivering a working repository test different abilities. To deliver a repository, an agent must turn a product-level description into an architecture, public interfaces, dependencies, and a build configuration, then find and fix integration failures from terminal output over a long session. Widely used evaluations do not require this. They either isolate a short program~\citep{chen2021codex,austin2021mbpp} or start from a mature repository and ask for a local patch~\citep{jimenez2024swebench}, so the architecture, dependency graph, packaging, and build system already exist.

Recent benchmarks move toward from-scratch development but draw the task boundary differently. Commit0 supplies library scaffolds and interactive test feedback~\citep{zhao2024commit0}. NL2Repo-Bench starts from an empty workspace for Python projects and restores reference packaging and tests before evaluation~\citep{ding2025nl2repo}. DevBench evaluates stages of software development as separate tasks~\citep{li2024devbench}, RepoGenesis targets Python and Java web microservices~\citep{peng2026repogenesis}, and RepoZero frames repository generation as cross-language behavioral reproduction from API specifications~\citep{zhang2026repozero}. None of these settings asks whether an agent can deliver a complete repository in its native ecosystem, including its own packaging and build artifacts, without seeing the acceptance tests while it works. Answering that question requires the agent to own the entire deliverable, the acceptance tests to stay sealed, reuse of the upstream implementation to be blocked, and specifications, environments, and tests to remain consistent across ecosystems.

\begin{figure}[t]
  \centering
  \includegraphics[width=\linewidth]{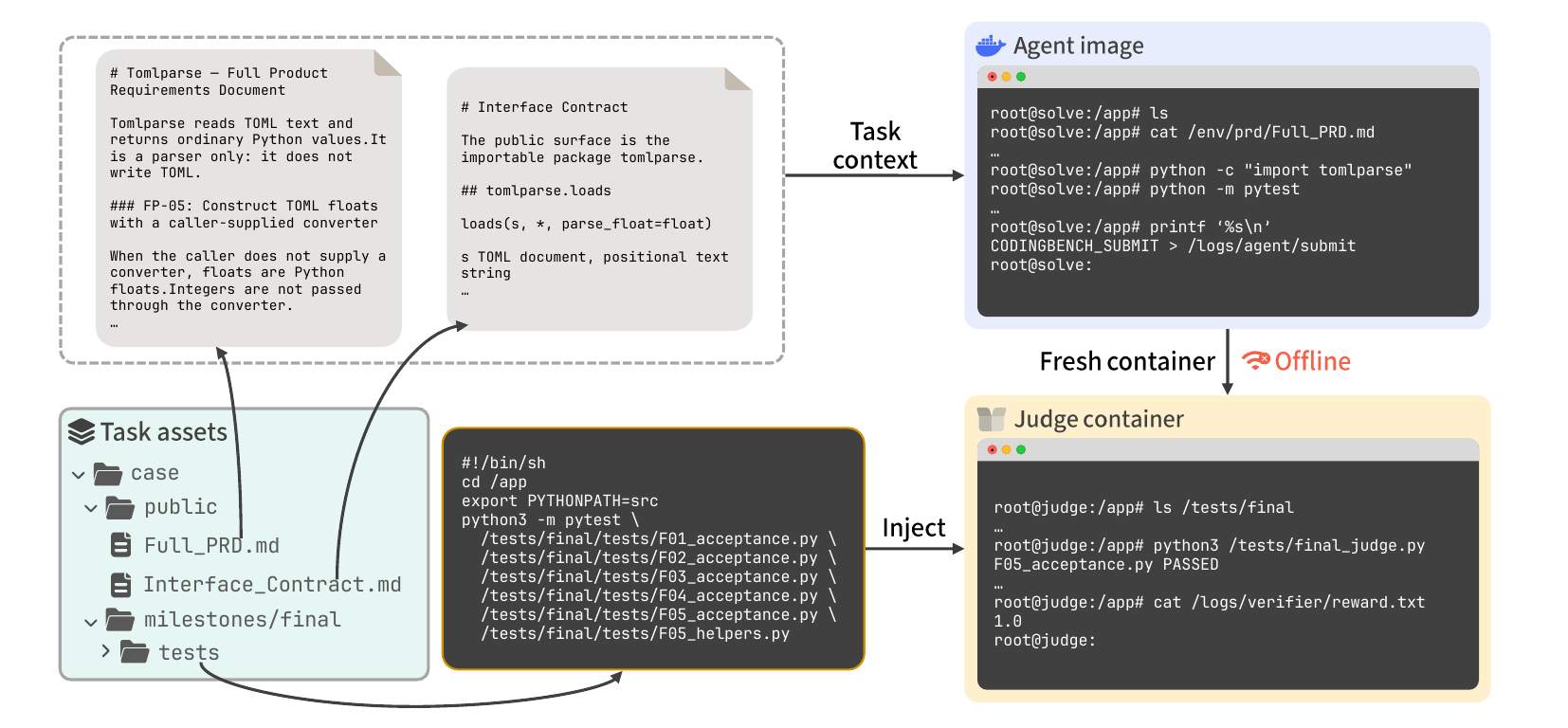}
  \caption{Data flow for one \bench task (Tomlparse). The \prd and \contract from the task assets are given to the agent as task context in the agent image, where it builds the repository in \texttt{/app} and submits by writing the submit file. The workspace is then copied into a fresh, offline judge container, the hidden acceptance tests are injected, and the judge writes a binary reward.}
  \label{fig:intro}
\end{figure}

\bench is built around this requirement.\footnote{Code, released tasks, and experiment scripts: \url{https://github.com/Gradient-Data/Zero2Repo}. Homepage: \url{https://zero2repo.ai/}.} Each task provides a product requirements document (\prd), an \contract, a reproducible toolchain, and an empty workspace. The \prd describes the required behavior; the \contract fixes the observable API and command-line surface and leaves architecture and implementation to the agent. The agent creates the repository, installs dependencies, builds the software when required, and submits the workspace. A task is solved only if every hidden acceptance test passes, and no LLM judge contributes to the score. Figure~\ref{fig:intro} traces one task through this process.

 The current release contains 17 tasks: nine in Python, one of which also includes Node.js hooks, two each in TypeScript, Go, and C, and one each in C++ and Zig. They cover parsers, protocols, command-line tools, security utilities, navigation filters, and software that depends on external resources. We plan to extend the benchmark to 30--50 tasks with broader language coverage for the v1 release.

Tasks are derived from real, version-pinned open-source projects by an automated pipeline in which executable evidence decides what is released. Language models read code and draft artifacts; each requirement must correspond to runnable behavior of the pinned project, each environment must rebuild from its recipe, and each test suite must pass a reference implementation and survive adversarial validation. Public names and repository identifiers are neutralized before release. During evaluation, the agent's image contains no hidden tests, code-hosting sites are unreachable, and installing or importing the upstream implementation is blocked. After an explicit submission, the workspace is copied into a fresh judge container where the hidden tests are injected and run. The judge neither repairs the repository nor supplies missing components.

We evaluate 18 model--harness configurations on the 17-task release, one adopted high-effort run per task. The first 11 tasks were deliberately drawn from well-known projects that models have very likely encountered in training. Even so, no configuration solves every task. The strongest solves 15 of 17, and both of its failures pass at least 96\% of the hidden tests. Across all configurations, most judged failures are near misses, but 51 unsuccessful runs never receive a complete judge result, and one task is solved by none. Our contributions are:
\begin{itemize}[leftmargin=*,itemsep=2pt]
  \item \textbf{A spec-to-repository benchmark in native ecosystems.} Agents deliver complete repositories from an empty workspace; acceptance is sealed, deterministic, and all-or-nothing. The design is language-agnostic, and the current release covers Python, TypeScript, Go, C++, C, and Zig.
  \item \textbf{An evidence-gated construction method with little human effort.} Real pinned repositories are turned into neutralized specifications, reproducible environments, and implementation-independent tests. Every task is grounded in the executable behavior of a real project and checked by reference execution and adversarial validation; human review is limited to inspecting the resulting artifacts.
  \item \textbf{A controlled evaluation with interpretable failures.} Under a common submission and judging protocol, the benchmark separates 18 configurations and records whether a zero is a judged near miss, a timeout, or a missing submission. One task remains unsolved by every configuration.
  \item \textbf{Reusable training assets.} The same task assets yield supervised trajectories and reinforcement-learning environments with the hidden suite as a verifiable reward. The outcome split in \S\ref{sec:analysis} indicates which failures are missing coverage on a judged submission and which are runs that never reach the judge.
\end{itemize}

\section{Related Work}
\label{sec:related}

\paragraph{Local code generation and repository modification.}
HumanEval, MBPP, LiveCodeBench, and BigCodeBench evaluate short programs~\citep{chen2021codex,austin2021mbpp,jain2024livecodebench,zhuo2024bigcodebench}; RepoBench and CrossCodeEval add repository context but keep completion as the target~\citep{liu2023repobench,ding2023crosscodeeval}. The SWE-bench family evaluates issue resolution in existing repositories with the project's own tests~\citep{jimenez2024swebench,openai2024swebenchverified,zan2025multiswebench}. In all of these settings the architecture, dependency graph, and build system already exist.

\paragraph{From-scratch software construction.}
Benchmarks for new repositories differ in their starting artifacts and success criteria (Table~\ref{tab:benchmark_comparison}). DevBench scores individual development stages with executable checks and an LLM judge~\citep{li2024devbench}; Commit0 reconstructs Python libraries from skeletons with interactive test feedback~\citep{zhao2024commit0}; NL2Repo-Bench starts from an annotator-written specification and an empty workspace, but restores reference packaging and tests before running the upstream pytest suite, which limits it to Python~\citep{ding2025nl2repo}; RepoGenesis targets Python and Java microservices~\citep{peng2026repogenesis}; RepoZero reformulates generation as cross-language reproduction checked by output equivalence~\citep{zhang2026repozero}. \bench starts from a behavioral product specification, requires a repository in the project's native ecosystem with its own packaging and build artifacts, seals all acceptance tests during the solve, and awards credit only for a complete pass.

\begin{table*}[t]
\caption{Benchmarks closest to \bench.}
\label{tab:benchmark_comparison}
\centering
\scriptsize
\setlength{\tabcolsep}{3pt}
\renewcommand{\arraystretch}{0.95}
\newcolumntype{L}[1]{>{\raggedright\arraybackslash}m{#1}}
\rowcolors{2}{blue!7}{white}
\begin{tabular}{L{1.9cm}L{3.1cm}L{2.7cm}L{2.0cm}L{3.2cm}}
\toprule
\textbf{Benchmark} & \textbf{Initial artifact} & \textbf{Required deliverable} & \textbf{Languages} & \textbf{Task construction} \\
\midrule
\multicolumn{5}{l}{\emph{Long-horizon agent tasks}} \\
SWE-Together & Repository and first user turn & Repository changes & Mixed & 109 tasks from recorded sessions \\
TerminalWorld & Container and instruction & Completed terminal task & Mixed & Derived from command recordings \\
FrontierSWE & Prepared environment & Open-ended solution & Mixed & 17 expert-curated tasks \\
\midrule
\multicolumn{5}{l}{\emph{From-scratch repository construction}} \\
DevBench & Stage-dependent repository context & Per-stage artifacts & Python, Java, JS, C/C++ & 22 curated repositories \\
Commit0 & API specification and skeleton & Library implementation & Python & 54 curated libraries \\
NL2Repo-Bench & Specification; empty workspace & Repository implementation & Python & 104 annotator-written specifications \\
RepoGenesis & Requirements and dependency file & Deployable web microservice & Python, Java & 106 repositories, 30 evaluated \\
RepoZero & API specification; empty workspace & Cross-language reproduction & Cross-language & Automated from open-source repositories \\
\bench & \prd and \contract; empty workspace & Complete native-ecosystem repository & Mainstream languages & Model-assisted derivation with evidence gates \\
\bottomrule
\end{tabular}
\end{table*}

\paragraph{Long-horizon agent evaluation and task construction.}
Terminal-Bench evaluates agents on difficult command-line tasks in containerized environments, pairing each task with a human-written solution and executable tests~\citep{terminalbench2025}. FrontierSWE poses open-ended engineering and research problems with 20-hour budgets and hidden graders; its 17 tasks are curated by domain experts, which makes the suite costly to extend~\citep{proximal2026frontierswe}. SWE-Together reconstructs multi-turn tasks from recorded user sessions and replays user feedback through an LLM simulator; whether a task can be solved therefore depends on the simulator conveying the original intent correctly~\citep{wu2026swetogether}. SetUpAgent and TerminalWorld automate task construction from historical execution environments and real command recordings, respectively~\citep{vergopoulos2025setupagent,chu2026terminalworld}. \bench targets repository delivery without prescribing a solution path, and it automatically authors the product specification, public interface, native build environment, and acceptance tests of each task, releasing a task only after reference execution and adversarial validation confirm that its tests are both passable and discriminating.

\section{The \bench Benchmark}
\label{sec:benchmark}

\subsection{Task Formulation}
\label{sec:task}
\label{sec:task:setting}

\begin{figure}[t]
  \centering
  \includegraphics[width=\linewidth]{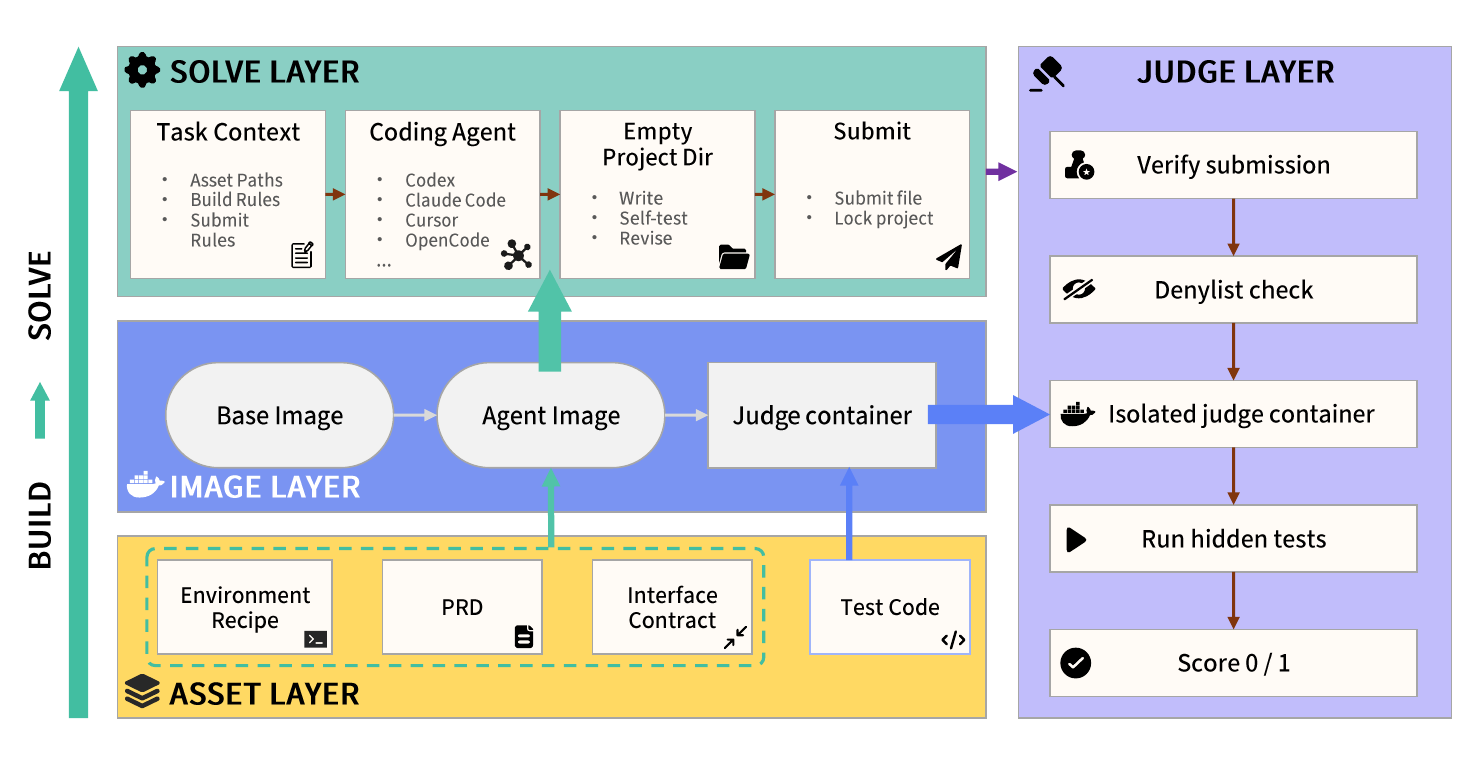}
  \caption{\bench evaluation architecture. \emph{Asset layer}: the environment recipe, \prd, and \contract are built into the agent image; the test code is used only by the judge. \emph{Image layer}: a base image shared by all tasks, a per-task agent image built from the environment recipe without hidden tests, and the judge environment, which is a fresh container started from the agent image with the hidden tests injected rather than a separately built image. \emph{Solve layer}: the agent receives the task context (specification paths, build rules, and submission rules), works in an empty project directory, and submits by writing a submit file. \emph{Judge layer}: the submission is verified and checked against the package denylist, the workspace is copied into the isolated judge container, the hidden tests are run, and a binary score is recorded.}
  \label{fig:overview}
\end{figure}

A \bench task is a tuple $(\mathcal{S},\mathcal{E},\mathcal{T})$ of a public specification $\mathcal{S}$, an execution environment $\mathcal{E}$, and a hidden acceptance suite $\mathcal{T}$. The agent receives $\mathcal{S}$ and an empty workspace inside $\mathcal{E}$, in which the required toolchain and dependencies are already installed. Its objective is to deliver a working repository that implements the specified behavior and interfaces (Figure~\ref{fig:overview}).

\paragraph{Public specification and implementation freedom.}
\label{sec:task:inputs}
The public specification states what the repository must do and how its functionality is accessed. The \prd groups behavioral requirements into feature points, each covering normal operation, boundary conditions, and failure cases. The \contract fixes the package names, API and CLI entry points, argument conventions, return types, and required error types through which those requirements are tested. The task also declares its language, build command, and hardware requirements. Within these constraints the agent chooses its own architecture, algorithms, and code organization.

\paragraph{Development and deliverable.}
The agent develops and tests the repository within a wall-clock budget $B$, without access to the hidden tests, a reference implementation, or acceptance feedback. It submits a workspace $W$ containing the source code, project configuration, and any build artifacts required by the public build command. The submission protocol is described in \S\ref{sec:harness}.

\paragraph{Success criterion.}
Each valid submission $W$ is scored by running the hidden acceptance suite $\mathcal{T}$ in environment $\mathcal{E}$:
\[
  R(W) = \mathbf{1}\!\left[
    W \text{ passes all tests in } \mathcal{T}
    \text{ under } \mathcal{E}
  \right].
\]
The benchmark score is the mean of $R$ over tasks (\S\ref{sec:experiments:setup}).

\paragraph{Example requirement.}
In the Tomlparse task, one \prd requirement is to support a caller-supplied floating-point converter without changing integer parsing. The \contract exposes this capability as \nolinkurl{tomlparse.loads(s, *, parse_float=float)}. The acceptance tests pass \texttt{Decimal} as the converter and check that parsing \texttt{0.1} yields \texttt{Decimal("0.1")} without intermediate binary-float rounding, while integer tokens bypass the converter.

\subsection{Dataset Construction and Verification}
\label{sec:authoring}

\begin{figure}[t]
  \centering
  \includegraphics[width=\linewidth]{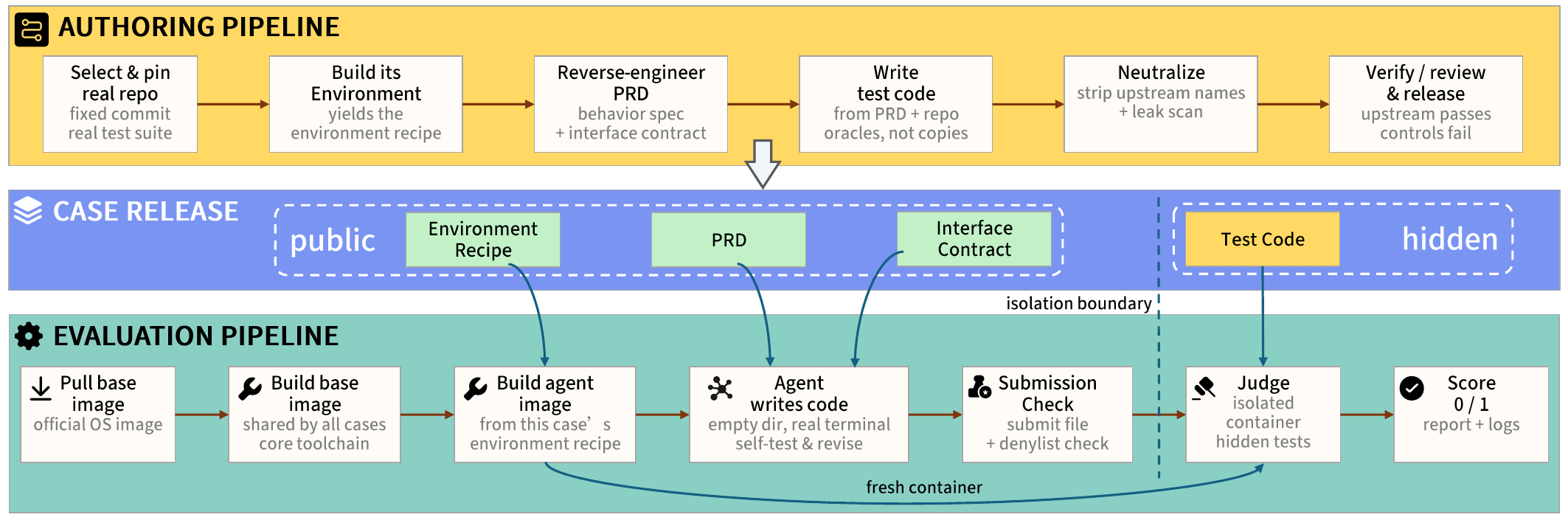}
  \caption{\bench pipeline. \emph{Authoring}: a repository is pinned, its environment recipe is built, a \prd and \contract are reverse-engineered from its behavior, test code is written from the \prd and the repository, public names are neutralized, and the task is verified, reviewed, and released. \emph{Release}: the environment recipe, \prd, and \contract are public; the test code is hidden. \emph{Evaluation}: the base image is pulled and built once, the agent image is built from the task's environment recipe, the agent writes code in an empty directory, and the submission is checked. Across the isolation boundary, the judge runs the hidden tests in a fresh container started from the agent image and records a binary score with logs.}
  \label{fig:pipeline}
\end{figure}

Each task is derived from an open-source repository. Language models derive the specification and tests from the repository's source, documentation, and test suite; executable checks and review then decide whether the task is released (Figure~\ref{fig:pipeline}).

\paragraph{Source selection and specification generation.}
We select repositories with a working installation procedure, an existing test suite, and functionality reachable through public APIs or CLIs, and pin each to a source revision. The first 11 tasks were chosen deliberately from widely used, long-established projects that are very likely present in model training data. This choice tests whether agents can fully rebuild a repository they have plausibly seen; later releases add recently published repositories and balance the language distribution. From the pinned code, documentation, and tests, language models write the \prd feature points and the corresponding \contract entries. Implementation details are kept only when they define an interface or a build constraint. An environment recipe records the toolchain, dependencies, and installation and build commands; external resources needed at acceptance time are listed with download URLs and checksums.

\paragraph{Acceptance tests and neutralization.}
Acceptance tests turn \prd feature points into executable checks against the interfaces fixed by the \contract. Upstream tests are rewritten against these interfaces so that they no longer depend on private helpers, source layout, or undocumented implementation choices. Expected outcomes are computed without calling the candidate, using standard-library computations, fixed vectors, or external processes; a signing test, for example, compares the output with an independently computed HMAC. Before release, identifying cues in public artifacts, such as project and author names and repository URLs, are replaced with neutral names, and interface renames are propagated to the tests. A per-task list of sensitive terms drives an identifier-boundary scan for residual cues. Neutralization removes explicit identity cues; it does not prevent a model from recognizing the project or recalling its code.

\paragraph{Verification and review.}
A task is released only after it passes execution checks and review. The environment must rebuild from its recipe, and a reference implementation adapted from the pinned source to the neutral interfaces must pass the full acceptance suite. Adversarial validation then checks that the suite rejects incorrect implementations: implementations with injected defects or prohibited shortcuts must receive zero reward. Structural checks verify that requirements, interface references, and test references agree. Model-assisted review and author inspection look for specification gaps, unintended structural constraints, and weak tests. Any defect found sends the affected artifacts back for revision and revalidation. Each released task went through such rounds until none found a defect, at most four per task, followed by author inspection (Appendix~\ref{app:controls}).

\paragraph{Dataset composition.}
\label{sec:task:suite}
The current release contains 17 tasks (Appendix Table~\ref{tab:cases}): parsers, security libraries, a protocol state machine, a CLI framework, a version-control large-file client, an NLP toolkit, a navigation filter, and a prose-pattern detector. Nine tasks are Python, one of them also including Node.js hooks, two each are TypeScript, Go, and C, and one each is C++ and Zig, so the release does not support balanced comparisons across languages. The tasks contain 138 feature points and 144 acceptance modules, between 4 and 18 per task. A module groups related test cases, so module counts are smaller than test-case counts.

\subsection{Evaluation Framework}
\label{sec:harness}

The \cbrun runner executes each trial in two phases, solve and judge, and records the same information for every configuration.

\paragraph{Agent execution and submission.}
\label{sec:harness:model}
\label{sec:harness:submit}
\label{sec:harness:denylist}
The agent runs in the task's solve container. The instruction names the paths of the \prd and \contract rather than inlining them, states the build contract, and explains how to submit (Appendix~\ref{app:instruction}). Code-hosting sites are blocked, package-manager shims refuse to install the upstream package, and a post-submission scan rejects imports of the upstream implementation. A submission is an explicit marker file checked after the agent exits; a successful CLI exit is not a submission. A denylist violation found by the scan allows one correction attempt, with no hidden-test feedback.

\paragraph{Independent acceptance.}
\label{sec:harness:isolation}
The agent image contains no hidden tests. After a valid submission, the runner copies the workspace into a fresh container started from the agent image with networking disabled, injects the hidden suite, and runs it. The copy keeps local dependencies, build artifacts, and agent-written tests; it drops changes outside the workspace, globally installed packages, and solve-time environment variables. The judge does not run the project's installation or build commands, so the workspace must already contain the artifacts named by the build contract; acceptance tests may compile their own test drivers. Judge output is stored for analysis and never returned to the agent.

\paragraph{Budgets and outcomes.}
The solve and judge phases each have a wall-clock limit of 7200 seconds. A task receives reward 1 only if the judge report shows that at least one test ran and every test passed; failures, errors, and missing submissions receive 0. Trial records distinguish solve timeouts, incomplete judge runs, and test failures, and separate infrastructure failures from implementation failures. Supported agent harnesses and recorded trial metadata are described in Appendix~\ref{app:versions}.

\section{Experiments}
\label{sec:experiments}

\subsection{Setup}
\label{sec:experiments:setup}

\paragraph{Configurations.}
We report the 18 model--harness configurations on the public leaderboard.\footnote{Homepage, retrieved 18 October 2026: \url{https://zero2repo.ai/}.} The harnesses are Codex, Claude Code, OpenCode, and Cursor. Four models are run in two harnesses, so those pairs are separate sittings of the same model rather than a controlled harness comparison. Each reported cell is one adopted run. Every configuration uses high reasoning effort. Solve and judge phases each have a 7200-second limit (\S\ref{sec:harness}).

\paragraph{Protocol and metrics.}
The primary metric is the pass rate, the fraction of the 17 tasks with reward 1. A run that never submits, reaches the solve limit, stops before submission, or has its hidden tests interrupted scores 0. We also report the total agent time, the token total, and the API spend published for each configuration. Per-task pass and failure fractions are taken from the same leaderboard. Costs for Step 5 Preview, Seed 2.1 Turbo, and Hy4 Preview are the OpenRouter list price stated there; Hy4 per-task cells are displayed as \$0. Gemini 4 Argon is priced at \$2, \$0.10, and \$10 per million input, cached-read, and output tokens. DeepSeek V4.1 Flash is priced at \$0.08, \$0.008, and \$0.50 on the same three buckets. For Qwen3.8 Max, GLM-5.3, Step 5 Preview, and Seed 2.1 Turbo, the published configuration total does not equal the sum of the per-task dollar figures, and for the last two it also does not equal the sum of the per-task token figures. Table~\ref{tab:main_results} uses the published totals.

\subsection{Main Results}
\label{sec:experiments:main}

\begin{table}[t]
\caption{Results on the 17-task release, one adopted run per task. Every run uses high reasoning effort. Time is total agent time, and tokens and cost are the configuration totals published with the leaderboard. A task is solved only when every hidden test passes.}
\label{tab:main_results}
\begin{center}
\small
\setlength{\tabcolsep}{5pt}
\begin{tabular}{llrrrrr}
\toprule
\textbf{Model} & \textbf{Harness} & \textbf{Solved} & \textbf{Pass rate} & \textbf{Time} & \textbf{Tokens} & \textbf{Cost (\$)} \\
\midrule
GPT-6 Astra            & Codex       & 15/17 & 88.2\% & 4h 13m  & 25.1M  & 80.57 \\
Claude Opus 5.5        & OpenCode    & 14/17 & 82.4\% & 9h 35m  & 146.5M & 195.69 \\
GPT-6 Astra            & OpenCode    & 12/17 & 70.6\% & 5h 18m  & 25.8M  & 80.90 \\
Claude Opus 5.5        & Claude Code & 11/17 & 64.7\% & 10h 14m & 180.6M & 178.33 \\
Claude Sonnet 5.5      & Claude Code & 10/17 & 58.8\% & 6h 20m  & 67.1M  & 58.54 \\
Claude Sonnet 5.5      & OpenCode    &  9/17 & 52.9\% & 5h 7m   & 91.8M  & 71.07 \\
GPT-6 Sol              & Codex       &  9/17 & 52.9\% & 3h 45m  & 39.7M  & 16.48 \\
Gemini 4 Argon         & OpenCode    &  9/17 & 52.9\% & 13h 25m & 84.5M  & 85.41 \\
Grok 4.7               & Cursor      &  8/17 & 47.1\% & 15h 20m & 146.1M & 78.99 \\
DeepSeek V4.1 Flash (1017) & OpenCode &  7/17 & 41.2\% & 4h 13m  & 231.9M & 3.36 \\
GPT-6 Sol              & OpenCode    &  6/17 & 35.3\% & 5h 5m   & 38.5M  & 15.06 \\
Qwen3.8 Max            & OpenCode    &  6/17 & 35.3\% & 29h 21m & 218.2M & 25.74 \\
Kimi K3                & OpenCode    &  5/17 & 29.4\% & 13h 35m & 101.4M & 84.59 \\
GLM-5.3                & OpenCode    &  5/17 & 29.4\% & 21h 29m & 469.4M & 13.43 \\
Hy4 Preview            & OpenCode    &  4/17 & 23.5\% & 22h 10m & 290.4M & 24.29 \\
DeepSeek V4 Pro        & OpenCode    &  2/17 & 11.8\% & 16h 40m & 333.5M & 75.38 \\
Step 5 Preview         & OpenCode    &  1/17 &  5.9\% & 15h 55m & 214.1M & 18.53 \\
Seed 2.1 Turbo         & OpenCode    &  1/17 &  5.9\% & 11h 0m  & 183.0M & 6.87 \\
\bottomrule
\end{tabular}
\end{center}
\end{table}

Table~\ref{tab:main_results} summarizes the results. Figure~\ref{fig:pass_heat} shows the percentage of hidden tests passed on each of the same runs. GPT-6 Astra with Codex solves 15 of 17 tasks. Claude Opus 5.5 with OpenCode solves 14. The same models in the other harness solve 12 and 11. Pass rates then fall to one task out of 17 for Step 5 Preview and Seed 2.1 Turbo. No configuration solves every task.

\FloatBarrier
\begin{figure}[p]
  \centering
  \input{figures/pass_rate_heat}
  \caption{Percentage of hidden tests passed by each adopted run. Rows follow Table~\ref{tab:main_results}. Columns are tasks 001--017. A cell is filled with the judge-layer background of Figure~\ref{fig:overview} when every hidden test passed, and is white otherwise. Numbers are that percentage rounded to one decimal place. A cell without a number was not judged; the symbols under the grid give the reason.}
  \label{fig:pass_heat}
\end{figure}

The two failures of the leading configuration are both judged submissions: task 003 passes 505 of 507 hidden tests, and task 014 passes 308 of 320. Claude Opus 5.5 with OpenCode fails tasks 003, 014, and 016, at 505/507, 302/320, and 387/388. Every failure of these two configurations therefore passes at least 94\% of the hidden tests. That pattern does not hold for the full suite. Of 306 runs, 134 pass and 172 do not. The 172 unsuccessful runs comprise 121 judged failures, 28 runs that hit the two-hour limit, 19 runs that exited without a submission, three runs stopped because reasoning hit the token limit, and one run whose hidden tests were interrupted. Of the 121 judged failures, 95 pass at least 90\% of their hidden tests, 80 pass at least 95\%, and 23 pass at least 99\%. Twenty-six pass less than 90\%, and three pass less than half: task 009 under GPT-6 Sol with Codex (150/498), task 015 under DeepSeek V4 Pro (73/169), and task 014 under GPT-6 Sol with Codex (151/320).

\paragraph{Time and cost.}
Longer and more expensive runs are not the runs that solve more tasks. The leading configuration uses 4 hours 13 minutes and \$80.57. The second-place configuration uses 9 hours 35 minutes and \$195.69, the highest spend in the table. GPT-6 Sol with Codex solves 9 of 17 tasks in 3 hours 45 minutes for \$16.48. DeepSeek V4.1 Flash solves 7 for \$3.36. Qwen3.8 Max records the longest total, 29 hours 21 minutes, and solves 6.

\paragraph{The harness changes the result.}
Holding the model fixed, the harness changes the pass count by three tasks for GPT-6 Astra (15 against 12), Claude Opus 5.5 (14 against 11), and GPT-6 Sol (9 against 6), and by one task for Claude Sonnet 5.5 (10 against 9). The disagreements are not a uniform shift. GPT-6 Astra with OpenCode solves task 003, which the Codex run misses, and misses tasks 006, 009, 010, and 016, which the Codex run solves. These are separate adopted runs, so the gap mixes harness behavior with run-to-run variation. It is large enough that a model name without the harness does not identify a result.

\paragraph{Results by task.}
\label{sec:experiments:language}
Appendix Table~\ref{tab:task_difficulty} counts how many of the 18 configurations solve each task. Task 004 is solved by 17, task 014 by none. Task 016 is solved only by GPT-6 Astra with Codex. Task 006 is solved by two configurations, and tasks 003, 009, and 010 by three. The leading configuration solves every task whose leaderboard language is Python, including task 016, which also contains Node.js hooks. It also solves both Go tasks, the C++ task, and the Zig task. It solves one of the two TypeScript tasks and one of the two C tasks, missing task 003 and task 014. Because several languages are represented by one or two tasks, these counts describe those tasks. They do not support a ranking of languages.

\section{Failure Analysis}
\label{sec:analysis}
\label{sec:analysis:failures}

The pass rates in Table~\ref{tab:main_results} say which configurations deliver a repository. The per-task fractions in Figure~\ref{fig:pass_heat} say what an unsuccessful run actually produced. We use the 306 adopted runs on the public leaderboard. This section does not assign individual failed tests to root causes. That attribution has to be read from the judge log of each adopted submission, and the submission-level fractions are what the leaderboard establishes.

\paragraph{A judged failure is usually a near miss.}
Of the 121 runs that submitted and were judged short of a full pass, 95 pass at least 90\% of their hidden tests and 80 pass at least 95\%. The two strongest configurations fail only in this band: their five unsuccessful tasks pass at least 94\% of the hidden tests. Under all-or-nothing acceptance, task 016 can fail at 387 of 388 tests, and task 003 at 505 of 507. A missing rule is enough. The same table also shows that a near miss is not the only way to fail. Twenty-six judged submissions pass less than 90\% of their tests, and on task 014 the judged scores range from 151/320 to 316/320. Task 014 is the only task solved by no configuration. The strongest submissions are close, and several weaker ones are not.

\paragraph{Many unsuccessful runs are never judged.}
Fifty-one of the 172 unsuccessful runs produce no complete judge result. Twenty-eight hit the two-hour solve limit, 19 exit without writing a submission, three stop because reasoning hits the token limit, and one has its hidden tests interrupted. These runs score 0 for a different reason than a submission that misses a rule. On task 014, five of the 18 configurations stop at the time limit or never submit, so their score does not measure how much of the navigation filter they implemented. Treating every zero as a near miss would erase that distinction. The binary reward still does not award partial credit. The recorded outcome does separate a near miss from a run that never reached the judge.

\paragraph{The hard tasks are not the same failure.}
Task 014 is unsolved, but the next-hardest tasks fail in different places. Task 016, a prose-pattern detector with Node.js hooks, is solved only by GPT-6 Astra with Codex. Task 006, the CLI framework, is solved by that configuration and by Claude Opus 5.5 with OpenCode. Tasks 003, 009, and 010 are solved by three configurations each, and not by the same three. Task 003 is solved by GPT-6 Astra with OpenCode and by both Claude Sonnet 5.5 sittings, not by GPT-6 Astra with Codex. Task 009 is solved by GPT-6 Astra with Codex and by both Claude Opus 5.5 sittings. A single unsolved-task list would hide that split.

\paragraph{Implications for agents.}
When a run is judged, the usual remaining gap is a small part of a suite the agent otherwise passes. The agents can check that gap before submission: the public specification states the behavior, and the hidden tests are not required to notice that 387 of 388 is not a pass. The 51 runs that never receive a complete judge result are a different gap. A verification loop does not help a session that exits without a submission or spends the whole budget without one. The harness already treats a missing or invalid submit file as a failure and does not run the hidden tests (\S\ref{sec:harness}). The leaderboard shows that this outcome is common below the top of the table, not an edge case.

\paragraph{Implications for training.}
Two consequences follow from the outcome split, without a finer attribution of individual tests. Reinforcement learning that uses the full-repository all-pass result as the reward gives the same penalty to a missed rule and to a missing submission, which matches the benchmark score and does not match the two different defects above. Supervised data and diagnostic rewards can separate them: a submission at 99\% needs coverage of the remaining specified behavior, while a timeout or a missing submission needs the agent to finish and submit a workspace the judge can run. The task assets provide the hidden suite as a verifiable reward in either case (\S\ref{sec:task}). Per-test labels for the adopted submissions are not part of this release.

\section{Conclusion, Limitations, and Future Work}
\label{sec:conclusion}
\label{sec:analysis:limitations}

\bench measures whether a coding agent can take a product specification and an empty directory and deliver a repository that runs in its native ecosystem. Two results stand out. First, on 17 tasks and 18 model--harness configurations, the benchmark separates current systems: GPT-6 Astra with Codex solves 15, Claude Opus 5.5 with OpenCode solves 14, and the pass rate falls to 1 of 17 at the bottom of Table~\ref{tab:main_results}. The same model in a second harness moves the score by as many as three tasks. Judged failures are usually near misses, while 51 unsuccessful runs never receive a complete judge result, so a zero is not one kind of failure (\S\ref{sec:analysis}). Second, the authoring pipeline is language-agnostic, and its validation stages leave one to two hours of author inspection per task, so the benchmark can grow across languages and be refreshed with newly published repositories without manual authoring. The same task assets also yield supervised trajectories and reinforcement-learning environments.

\paragraph{Limitations.}
The main limitation is scale. The current release contains 17 tasks, nine of them Python, with two tasks each in TypeScript, Go, and C and one each in C++ and Zig. One of the Python tasks also includes Node.js hooks. Non-Python results still describe one or two tasks, and the first 11 seed projects are well known, so pass rates on that subset are probably higher than they would be on unseen repositories. Extension to 30--50 tasks with a more balanced language distribution and a mix of older and recently published repositories is in progress. The evaluation is also limited in coverage. Each configuration contributes one adopted run per task, so run-to-run variance is not measured, and the harness gaps in \S\ref{sec:experiments:main} are not a controlled comparison. Claude Fable 5.1 was excluded because it sometimes declined to proceed on safety grounds during task construction. Repeated runs, and a failure analysis that attributes each missed test in the adopted submissions, are planned for the next release.

\paragraph{Future work.}
The benchmark will be versioned: as model training data moves forward, new versions will draw tasks from the most recently published repositories, and where possible from private repositories, to limit exposure in training data.


\subsection*{AI use statement}

Generative AI tools were used in three ways. First, they are part of the task-authoring pipeline described in \S\ref{sec:authoring}: models read upstream source code, draft the product requirements documents and interface contracts, propose acceptance tests, and run commands during task construction. Every artifact produced this way passed the execution checks and adversarial validation described in \S\ref{sec:authoring} before release, and the released tasks were inspected by the authors. Second, we used generative AI tools to draft and revise sections of the manuscript, to translate drafts between Chinese and English, and to produce initial versions of the schematic figures, which the authors then redrew. Third, the coding agents evaluated in \S\ref{sec:experiments} are generative AI systems and are the object of study. All claims, numbers, and analyses in the paper were checked by the authors against the experiment records, and the authors take full responsibility for the content of this work.

\subsection*{Ethics statement}

\bench is built from publicly available open-source repositories released under permissive licenses (for example MIT, BSD, and Apache-2.0), and derived artifacts follow the terms of the corresponding licenses. Public artifacts are neutralized to remove author and organization names and other identifying information; hidden tests and environment recipes are derived from the projects' own tests and build configurations. The benchmark evaluates software agents, not human subjects. Releasing a benchmark on which agents can be trained may accelerate saturation; the benchmark is therefore versioned, and new versions draw tasks from recently published repositories. Denylists containing upstream package names are distributed privately so that the release does not publish a direct mapping from neutralized tasks to upstream projects.

\subsection*{Reproducibility statement}

The evaluation harness, the released tasks (public specifications, hidden suites, and environment recipes), and the scripts used to run the experiments are available at \url{https://github.com/Gradient-Data/Zero2Repo}. The project homepage is \url{https://zero2repo.ai/}. The task-authoring pipeline is not released; \S\ref{sec:authoring} and Figure~\ref{fig:pipeline} describe its stages and the checks a task must pass before release. \S\ref{sec:harness} describes the isolation and scoring procedure; Appendix~\ref{app:versions} lists the models, harnesses, and budgets used in the reported runs; Appendix~\ref{app:controls} describes the validation each task passed before release. Environments rebuild from public base images and recipe locks without a private cache. Trial outputs include the reward, judge logs, agent logs, and reproducibility metadata (agent-specification hash, model, CLI version, and the names of forwarded credentials), as described in Appendix~\ref{app:versions}.

\bibliography{refs}

\appendix
\section{Released Tasks}
\label{app:tasks}

\begin{table}[h]
\caption{Tasks in the current \bench release. The product name is the neutralized name given to the agent; the source is the upstream project the task was derived from. An acceptance module groups related test cases.}
\label{tab:cases}
\begin{center}
\scriptsize
\setlength{\tabcolsep}{4pt}
\begin{tabular}{lllllr}
\toprule
\textbf{Task} & \textbf{Product} & \textbf{Source} & \textbf{Language} & \textbf{Domain} & \textbf{\shortstack{Acceptance\\modules}} \\
\midrule
001 & Tomlparse      & tomli              & Python         & TOML v1.1 parser                       & 5  \\
002 & python-envfile & python-dotenv      & Python         & \texttt{.env} loading and expansion    & 8  \\
003 & ymlcodec       & js-yaml            & TypeScript     & YAML codec with schemas and tags       & 7  \\
004 & Signtoken      & itsdangerous       & Python         & HMAC signing and serialization         & 4  \\
005 & httpwire       & h11                & Python         & HTTP/1.1 protocol state machine        & 7  \\
006 & Optlyn         & Click              & Python         & CLI framework with shell completion    & 14 \\
007 & Otpkit         & pyotp              & Python         & One-time password generation           & 5  \\
008 & Hrefparse      & ada-url            & C++            & WHATWG URL parsing                     & 4  \\
009 & Git Orbulk     & git-lfs            & Go             & Version-control large-file client      & 18 \\
010 & Lingora        & NLTK               & Python         & NLP toolkit with model resources       & 7  \\
011 & PathSel        & jmespath           & Python         & JSON path query language               & 8  \\
012 & lint-policy    & anti-slop          & TypeScript     & Oxlint ruleset                         & 8  \\
013 & membundle      & okf-agent-memory   & Go             & Git-native agent memory                & 11 \\
014 & NAVFILTER      & INSLIB             & C              & 3D navigation filters                  & 11 \\
015 & oggrepack      & balrogg            & C              & Lossless Ogg recompressor              & 10 \\
016 & prosecheck     & sloptrim           & Python, Node   & Prose-pattern detector                 & 7  \\
017 & klyvmap        & zroar              & Zig            & 64-bit roaring bitmaps                 & 10 \\
\bottomrule
\end{tabular}
\end{center}
\end{table}

\section{Task Package Layout}
\label{app:layout}

Each task under \texttt{cases/<task\_id>/} contains the following files. Entries marked ``no'' are never present in the agent image.

\begin{center}
\resizebox{\linewidth}{!}{%
\begin{tabular}{llc}
\toprule
\textbf{Path} & \textbf{Role} & \textbf{Visible to agent} \\
\midrule
\texttt{public/Full\_PRD.md} & Product requirements & yes \\
\texttt{public/Interface\_Contract.md} & Public API / CLI contract & yes \\
\texttt{public/Hardware\_Requirements.md} & Optional hardware constraints & yes \\
\texttt{source/manifest.json} & Runner metadata, sensitive terms, judge bans & no \\
\texttt{source/recipe.lock.json} & Environment recipe (base image, install, build) & no \\
\texttt{source/env/resources.json} & Declared external resources (URL, size, SHA-256) & no \\
\texttt{source/denylist.json} & Upstream package and import bans & no \\
\texttt{milestones/final/tests/} & Hidden acceptance tests & no \\
\texttt{milestones/final/test\_manifest.json} & Test inventory and command template & no \\
\texttt{controls/<name>/} & Negative-control workspaces with expected reward & no \\
\bottomrule
\end{tabular}}
\end{center}

\section{Agent Instruction}
\label{app:instruction}

The instruction below is the text given to every agent for a task without a build step. The build-contract section at the end changes with the task's build command and working directory.

\begin{lstlisting}
# Development Task

You are an autonomous software engineer. Build the complete project described
in the specification files listed below, from scratch, in your current working
directory. Implement every step of the development plan so that the finished
project fully satisfies the specification.

Read those files in full before you start writing code.

When you are done, the workspace must be in a state the hidden tests can use
directly. See the Build contract below for whether a build step is required
and where outputs must remain.

---

## Your environment

* Your workspace is `/app` (your current working directory).
  Build the project here. See the Build contract below for how the judge
  locates outputs.
* The full specification is in these files. Read them in full before you start:
  the PRD at `/environment/prd/Full_PRD.md` and the Interface Contract at
  `/environment/Interface_Contract.md`.
* Language runtimes and dependencies the project needs are already installed in
  this image. You have network access for your own model/tool calls.

## How you work

* You may develop freely: there is no limit on the number of steps, turns, edits
  or commands. The only limit is a wall-clock time budget for the whole session.
* **Implement from scratch.** Build the target system described in the PRD and
  Interface Contract yourself. Do not download, clone, vendor, copy, or install
  an existing upstream implementation of that target system (for example via
  `pip install`, `npm install`, or `cargo add` for the product you are building).
  General-purpose libraries and tools are allowed; the described product behavior
  must be your own code.
* Do not treat a runtime- or toolchain-bundled implementation of the same kind
  of product as a dependency or an oracle. The behavior you deliver must be
  your own code.
* GitHub and other code-hosting sites for upstream projects are **not reachable**
  from this environment during your session.
* You are encouraged to write and run your OWN tests and checks repeatedly to
  validate your implementation against the PRD and Interface Contract, then fix
  and iterate. A real develop -> test -> debug loop is expected, not a single
  pass.
* When you are confident the implementation is complete, write the submit file
  described below, then you may end your session. Ending the session is **not**
  a submission.
* Submit by writing exactly this one-line file (no extra words):
  path `/logs/agent/submit`
  contents `CODINGBENCH_SUBMIT`
* If that file is missing or its contents are wrong, this attempt fails and the
  hidden acceptance tests will not run.

## How you are scored

* After a valid submit file is present, a hidden acceptance test suite is run
  against your final workspace state and produces a binary pass/fail reward.
* The hidden tests are NOT present in this container and you cannot access them.
  Do not look for them, and do not special-case any test: implement the public
  Interface Contract behavior fully and correctly.

## Build contract

* The judge does **not** run any install or build step for you. The hidden
  acceptance tests run against your final `/app` exactly as you leave it.
* There is no build step for this project.
* The hidden tests run against `/app` as the working directory.
\end{lstlisting}

\section{Example Specification Excerpt}
\label{app:example}

The following excerpts from task 001 (Tomlparse) show the level of abstraction of the \prd and the \contract. Text is quoted verbatim; omissions are marked with an ellipsis.

\paragraph{\prd, terminology (excerpt).}
\begin{center}
\small
\begin{tabular}{p{0.24\linewidth}p{0.68\linewidth}}
\toprule
\textbf{Term} & \textbf{Meaning in this PRD} \\
\midrule
TOML document & A Unicode text that the caller treats as one TOML v1.1.0 document. The product parses one document per call. \\
String-parse entry & The library entry that accepts TOML as a Python text string and returns a mapping, or fails. \\
Float converter & An optional callable the caller supplies so TOML floats (including \texttt{inf} and \texttt{nan} spellings) are built as something other than a Python float. Specified in FP-05. \\
Decode error & The product's documented parse-failure exception. It is a kind of value error. Specified in FP-04. \\
Discrimination & An assertion's ability to distinguish a faithful implementation from a hollow, skipped, or proxy one. \\
\bottomrule
\end{tabular}
\end{center}

\paragraph{\prd, feature point FP-05 (excerpt).}
\begin{quote}\small
\textbf{FP-05: Construct TOML floats with a caller-supplied converter.}
\emph{Public entry:} The optional float converter on both the string-parse entry and the binary-file parse entry. When the caller does not supply a converter, FP-02's default (Python float) applies. This feature point does not change integers, strings, booleans, date-times, tables, or arrays.

\emph{Normal behavior:} For \texttt{precision-matters = 0.982492} and a converter that builds a standard-library decimal from the text it is given, the value of \texttt{precision-matters} is a decimal equal to the decimal of the characters \texttt{0.982492}, not a binary float of that magnitude. [\dots] Integers are not passed through the converter. In \texttt{a = 1} then \texttt{b = 1.0}, with a decimal converter, \texttt{a} remains a Python integer and \texttt{b} is a decimal.

\emph{Boundary / error behavior:} If the converter returns a dictionary or a list (including a subtype of either), the parse fails with a value error. [\dots] This is not a decode error: the observer can tell an illegal converter result from invalid TOML.

\emph{Verifiable oracle:} [\dots] \emph{Failure / absence:} the converter is ignored and values stay Python floats; integers are also converted; a converter that returns a dictionary or a list is accepted and the result is treated as a table or array; the illegal-converter failure is reported as a decode error.
\end{quote}

\paragraph{\contract, \texttt{tomlparse.loads} (excerpt).}
\begin{quote}\small
\texttt{loads(s, *, parse\_float=float)}

\texttt{s} --- the TOML document as a Python text string. Passed positionally. A bytes object, a boolean, a file object, or any other non-text value is refused with \texttt{TypeError}, not \texttt{TOMLDecodeError}.

\texttt{parse\_float} --- optional callable that receives the spelling of a TOML float (including \texttt{inf} and \texttt{nan} tokens) and returns the constructed value. Keyword-only. The default is the builtin \texttt{float}. Integers, strings, booleans, and date-times are not passed through this converter. If the converter returns a dictionary or a list (including a subtype of either), the parse fails with \texttt{ValueError}, not \texttt{TOMLDecodeError}.
\end{quote}

\section{Validation Before Release}
\label{app:controls}
\label{sec:experiments:controls}

Every released task went through repeated rounds of validation followed by author inspection, until a round found no defect. Among the first 11 tasks, none needed more than four rounds (Table~\ref{tab:validation_rounds}). In each round, a frontier coding agent solved the task under the release harness, and the resulting run was reviewed along three dimensions. \emph{Run integrity}: the judge collected and ran the full hidden suite, the recorded counts are consistent, and the judged workspace matches the agent's final submission. \emph{Specification, tests, and oracles}: every behavior the tests check is stated in the public specification, expected values are computed independently of the candidate, and the environment does not leak the upstream implementation. \emph{Adversarial validation}: the passing implementation is mutated to break specified behavior, replaced by an empty or hollow workspace, or combined with a prohibited shortcut, and each variant must be rejected. The upstream implementation adapted to the neutral interfaces serves as the reference and must pass the full suite.

These reviews found defects that single-pass checks miss, for example a suite that accepted an implementation violating a stated no-I/O constraint, a status-code rule without a rejecting assertion, a toolchain that bundled the upstream implementation and could serve as an oracle, and a missing denylist entry. Each defect was fixed in the affected artifacts and the task was validated again in the next round. After its final round and author inspection, each of the first 11 tasks was released with no open defects. Tasks 012--017 were released through the same execution checks. Their validation-round counts are not in Table~\ref{tab:validation_rounds}.

\begin{table}[h]
\caption{Validation rounds for the first 11 tasks before release.}
\label{tab:validation_rounds}
\begin{center}
\small
\setlength{\tabcolsep}{5pt}
\begin{tabular}{lccccccccccc}
\toprule
\textbf{Task} & 001 & 002 & 003 & 004 & 005 & 006 & 007 & 008 & 009 & 010 & 011 \\
\midrule
\textbf{Rounds} & 3 & 1 & 3 & 3 & 2 & 4 & 2 & 2 & 3 & 3 & 2 \\
\bottomrule
\end{tabular}
\end{center}
\end{table}

This validation process is now part of the authoring pipeline and runs automatically for every new task. Author inspection is still required, but it takes about one to two hours per task. As agents become more capable, or with stronger and more expensive agents as reviewers, we expect task production to become fully automatic.

\section{Agent Integration, Versions, and Budgets}
\label{app:versions}
\label{sec:harness:backends}
\label{sec:harness:cost}

The runner supports Codex, Claude Code, OpenCode, and Cursor through declarative agent specifications, and each evaluated system is a model--harness configuration. Every trial records task and test provenance, image identifiers, the agent-specification hash, the resolved model, the CLI version, the budgets, execution logs, and wall time, as well as token usage when the harness reports it.

The runs in Table~\ref{tab:main_results} use high reasoning effort. Solve and judge phases each have a 7200-second limit. The leaderboard does not publish the CLI build for each adopted run, so this paper does not list one. Codex does not accept a temperature, so the Codex runs have none.

\section{Additional Results}
\label{app:results}

\begin{table}[h]
\caption{Tasks solved by the 18 configurations in Table~\ref{tab:main_results}. Hidden tests is the size of the suite on the adopted runs. Solved is the number of configurations with reward 1.}
\label{tab:task_difficulty}
\begin{center}
\small
\setlength{\tabcolsep}{6pt}
\begin{tabular}{lllr r}
\toprule
\textbf{Task} & \textbf{Product} & \textbf{Language} & \textbf{Hidden tests} & \textbf{Solved} \\
\midrule
001 & Tomlparse      & Python       & 302 & 14/18 \\
002 & python-envfile & Python       & 281 & 15/18 \\
003 & ymlcodec       & TypeScript   & 507 &  3/18 \\
004 & Signtoken      & Python       & 258 & 17/18 \\
005 & httpwire       & Python       & 395 &  9/18 \\
006 & Optlyn         & Python       & 476 &  2/18 \\
007 & Otpkit         & Python       &  80 & 16/18 \\
008 & Hrefparse      & C++          & 248 &  8/18 \\
009 & Git Orbulk     & Go           & 498 &  3/18 \\
010 & Lingora        & Python       & 197 &  3/18 \\
011 & PathSel        & Python       & 321 & 10/18 \\
012 & lint-policy    & TypeScript   & 532 &  8/18 \\
013 & membundle      & Go           & 585 &  7/18 \\
014 & NAVFILTER      & C            & 320 &  0/18 \\
015 & oggrepack      & C            & 169 &  5/18 \\
016 & prosecheck     & Python, Node & 388 &  1/18 \\
017 & klyvmap        & Zig          & 234 & 13/18 \\
\bottomrule
\end{tabular}
\end{center}
\end{table}

The by-task pass and failure fractions, token totals, and per-task spend are published at \url{https://zero2repo.ai/}. This paper does not repeat that matrix. Where a published configuration total does not equal the sum of the per-task figures, \S\ref{sec:experiments:setup} states which total Table~\ref{tab:main_results} uses.

\end{document}